# Realisation of de Gennes' Absolute Superconducting Switch with a Heavy Metal Interface

Hisakazu Matsuki[1], Alberto Hijano[2,3,4], Grzegorz P. Mazur[1,5], Stefan Ilić[2,4], Binbin Wang[6], Yuliya Alekhina[1], Kohei Ohnishi[7], Sachio Komori[8], Yang Li[1,9], Nadia Stelmashenko[1], Niladri Banerjee[10], Lesley F. Cohen[10], David W. McComb[6], F. Sebastián Bergeret[2,11], Guang Yang[1,12,13*] and Jason W. A. Robinson[1*]

1. Department of Materials Science & Metallurgy, University of Cambridge, 27 Charles Babbage Road, Cambridge CB3 0FS, U.K.
2. Centro de Física de Materiales (CFM-MPC) Centro Mixto CSIC-UPV/EHU, E-20018 Donostia-San Sebastián, Spain
3. Department of Condensed Matter Physics, University of the Basque Country UPV/EHU, 48080 Bilbao, Spain
4. Department of Physics and Nanoscience Center, University of Jyväskylä, P.O. Box 35 (YFL), Jyväskylä, FI-40014 Finland
5. QuTech and Kavli Institute of NanoScience, Delft University of Technology, 2600 GA Delft, The Netherlands
6. Department of Materials Science and Engineering, The Ohio State University, Columbus, OH 43210, USA.
7. Department of Electrical, Electronic and Communication Engineering, Kindai University, Osaka 577-8502, Japan
8. Department of Physics, Nagoya University, Nagoya 464-8602, Japan
9. Cambridge Graphene Centre, University of Cambridge, 9 JJ Thomson Avenue, Cambridge CB3 0FA, U.K.
10. Department of Physics, Blackett Laboratory, Imperial College London, London SW7 2AZ, U.K.
11. Donostia International Physics Center (DIPC), 20018 Donostia–San Sebastián, Spain
12. National Key Laboratory of Spintronics, Hangzhou International Innovation Institute, Beihang University, Hangzhou 311115, China
13. School of Integrated Circuit Science and Engineering, Beihang University, Beijing 100191, China

*e-mail: gy251@buaa.edu.cn, jjr33@cam.ac.uk

**In 1966, Pierre-Gilles de Gennes proposed a non-volatile mechanism for switching superconductivity on and off in a magnetic device. This involved a superconductor (S) sandwiched between ferromagnetic (F) insulators in which the net magnetic exchange field could be controlled through the magnetisation-orientation of the F layers. Because superconducting switches are attractive for a range of applications, extensive studies have been carried out on F/S/F structures. Although these have demonstrated a sensitivity of the superconducting critical temperature ($T_c$) to parallel (P) and antiparallel (AP) magnetisation-orientations of the F layers, corresponding shifts in $T_c$ (i.e., $\Delta T_c = T_{c,AP} - T_{c,P}$) are lower than predicted with $\Delta T_c$ only a small fraction of $T_{c,AP}$, precluding the development of applications. Here, we report EuS/Au/Nb/EuS structures where EuS is an insulating ferromagnet, Nb is a superconductor and Au is a heavy metal. For P magnetisations, the superconducting state in this structure is quenched down to the lowest measured temperature of 20 mK meaning that $\Delta T_c/T_{c,AP}$ is practically 1. The key to this so-called "absolute switching" effect is a sizable spin-mixing conductance at the EuS/Au interface which ensures a robust magnetic proximity effect, unlocking the potential of F/S/F switches for low power electronics.**

The original superconducting switch[1] modelled by de Gennes requires a thin-film superconductor (S) with a thickness ($d_s$) that is less than one superconducting coherence length ($\xi_s$), sandwiched between two ferromagnetic (F) insulators (Fig. 1**a** and 1 **b**). Due to the strong pair-breaking interaction between the S and F materials, the critical temperature ($T_c$) of the F/S/F structure is suppressed for a parallel (P) alignment of the magnetisation of the F layers. Conversely, if the magnetisation of the F layers aligns antiparallel (AP), the influence of the two F layers on the superconductivity cancels, in principle, meaning that the suppression of $T_c$ is reduced. An equivalent superconducting switch was later proposed by Tagirov[2] which involved transition metal ferromagnets (instead of ferromagnetic insulators), allowing superconductivity to penetrate the F layers causing an additional background suppression of $T_c$ in both the P and AP magnetic states. Both models predicted that for certain parameter combinations, not only should the $T_c$ difference between P and AP magnetic states [i.e., $\Delta T_c = T_{c,AP} - T_{c,P}$] be a significant fraction of $T_{c,AP}$, but also that superconductivity should be completely suppressed for all temperatures in the P-state – this is so-called "absolute switching" with $\Delta T_c/T_{c,AP} = 1$ meaning that F/S/F becomes a truly magnetically-controlled superconducting switch[3–8], a highly sought-after device for low power electronics.

The first[9] experimental demonstration of F/S/F switching was reported in 2002 with measured values of $\Delta T_c$ (roughly 6 mK) much lower than predicted[1,2]. In addition, because the temperature width of the superconducting transition $\sigma_{Tc}$ was larger than $\Delta T_c$, the resistance change at any temperature induced by the magnetic reorientation was small. Since then many papers have been published[10–22] using different materials combinations, largely focusing on transition metal ferromagnets in which the magnetic exchange field is dominated by spin-splitting of the *d*-orbitals and transport through hybridised *s-d* orbitals; however, values of $\Delta T_c$ are always dramatically lower than predicted by theory. Well-defined on and off switching of superconductivity has been demonstrated in limited F/S/F structures involving *f*-orbital magnets such as metallic Ho[19,23] or insulators including EuS[18] and GdN[20,24] with low $\sigma_{Tc}$, albeit over a narrow temperature range with $\Delta T_c/T_{c,AP} \ll 1$. However, the ultimate aim of absolute switching has not been achieved as yet to our knowledge.

Theoretically, absolute switching in a F/S/F structure requires a large proximity-induced magnetic exchange field ($h_{ex}$) in the S layer relative to its superconducting energy gap ($\Delta_0$) with a magnitude $h_{ex} > (\sqrt{2}/2)\Delta_0$ in the P-state[1]. It is well-established that $h_{ex}$ is proportional to the interfacial spin-mixing conductance $G_i$ (imaginary part) for constant $d_s$, and therefore, a large $h_{ex}$ corresponds to a large $G_i$[25–27]. Here, $G_i$ is a measure of the exchange field existing between the electrons in the non-magnetic metal and those in EuS and characterises the efficiency of F/N interfacial spin transport (see, e.g., Refs. 26,29,30). For an F/S interface, this leads to a spin-splitting of the superconducting density of states (Fig. 1**d**)[28]. Pioneering experiments on Al/EuS structures (where Al is a S layer) were performed by Meservey, Tedrow and Moodera[28,31–35]. They demonstrated a splitting in the superconducting density of states[28] that corresponded to a magnetic field of more than 1 T. Non-superconducting experiments on EuS/Pt[26] and EuS/Graphene[36] structures also show evidence for large proximity-induced exchange fields, larger than 10 T in both Pt and Graphene. We note the recent experiments on Nb/EuS wires showing a so-called supercurrent diode effect which can be related to a large $h_\text{ex}$ in the Nb[37] and/or vortices[38].

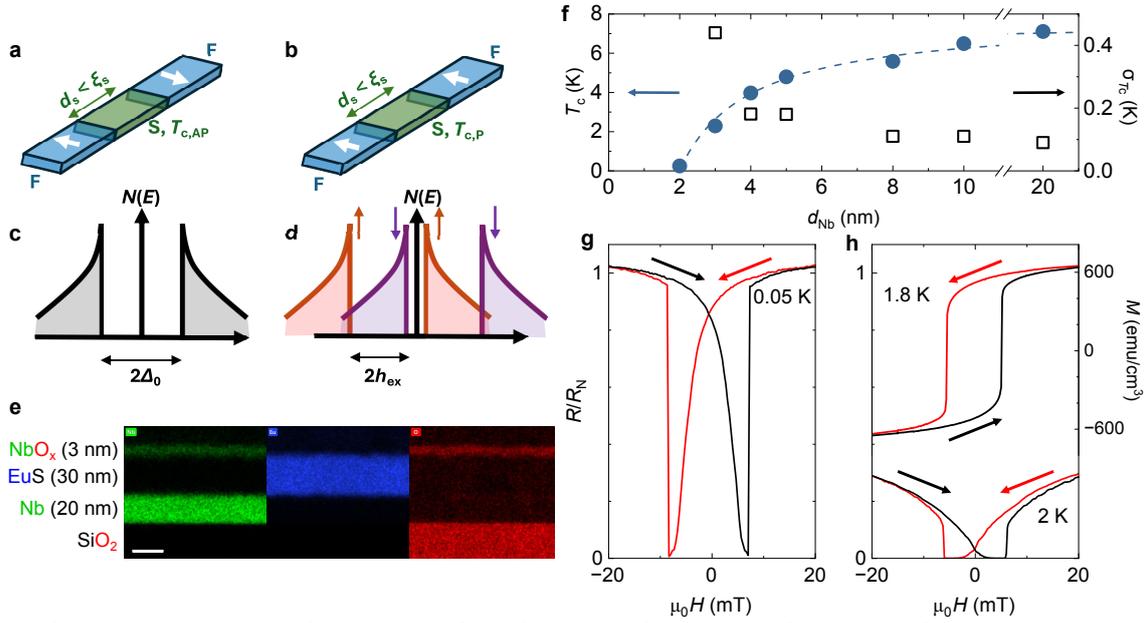

**Fig. 1: A de Gennes' superconducting switch and structural, superconducting, and magnetic properties of Nb(3 nm)/EuS(30 nm)/Nb($d_\text{Nb}$)/SiO$_2$//Si structures. a, b** Schematic diagrams of a F/S/F superconducting switch in which a superconductor (S) is sandwiched between ferromagnetic insulators (F): **a**, the proximity-induced magnetic exchange field ($h_\text{ex}$) in the S layer from the AP-aligned magnetisations is minimised or is, ideally, zero, preserving the superconducting state with a transition temperature $T_\text{c,AP}$; **b**, For P-aligned magnetisations, $h_\text{ex}$ is maximised so the superconducting transition temperature $T_\text{c,P}$ is much lower than $T_\text{c,AP}$. **c, d**, Representations of the superconducting density of states diagrams for the S layer for AP and P magnetisations of the F layers: **c**, in the AP-state the density of states shows no evidence of proximity-induced magnetism (i.e., $h_\text{ex}$ = 0) whereas in the P-state in **b** there is an energy splitting of $2h_\text{ex}$ in the spin-bands due to the proximity-induced exchange field. **e**, Chemistry diagram from a control sample of a Nb(3 nm)/EuS(30 nm)/Nb(20 nm)/SiO$_2$//Si structure showing Nb (green), Eu (blue) and O (red). The scale bar has a length corresponding to 20 nm. **f**, The left axis shows the zero-field-cooled superconducting transition temperature $T_\text{c}$ versus Nb thickness $d_\text{Nb}$ and the right axis shows the superconducting transition width $\sigma_{T_\text{c}}$ versus $d_\text{Nb}$. **g**, Normalised resistance $R$ versus in-plane magnetic field $H$ ($R(H)$) of an unpatterned Nb(3 nm)/EuS(30 nm)/Nb(2 nm)/SiO$_2$//Si structure at 50 mK, where $R_\text{N}$ is the normal state resistance. **h**, Normalised $R(H)$ of an unpatterned Nb(3 nm)/EuS(30 nm)/Nb(3 nm)/SiO$_2$//Si structure at 2 K along with the magnetisation vs in-plane magnetic field $M(H)$ hysteresis loop for a 30-nm-thick EuS film at 1.8 K. Red (black) curves indicate a decreasing (increasing) in-plane magnetic field.

Huertas-Hernando and Nazarov[39,40] theoretically proposed a modification of the F/S/F structures by inserting a normal metal layer (N) at the F/S interface as a means of achieving absolute switching. This N layer facilitates physical separation of the competing superconducting and magnetic order parameters and allows their careful control within N through superconducting and magnetic proximity effects. Here, we first report EuS/Nb/EuS structures with a superconducting switch efficiency $\Delta T_\text{c}/T_\text{c,AP}$ that can reach about 50%. In the next step, by inserting a 20-nm-thick heavy metal layer of Au at one interface (i.e., EuS/Au/Nb/EuS), we demonstrate a dramatic enhancement of $\Delta T_\text{c}/T_\text{c,AP}$ reaching 1, achieving absolute switching[1]. The key to the enhancement of $\Delta T_\text{c}/T_\text{c,AP}$ is related to the interface chemistry and a larger proximity magnetic exchange field in Au due to a large $G_\text{i}$ at EuS/Au interface versus EuS/Nb interface. These results are obtained in extremely thin layers of 4-nm-thick Nb in which the superconducting state is preserved in the AP-state with the P-state showing no evidence of superconductivity down to 20 mK.

A set of Nb(3 nm)/EuS(30 nm)/Nb($d_s$)/SiO$_2$//Si, Nb(3 nm)/EuS(20 nm)/Nb(4 nm)/EuS(10 nm)/SiO$_2$//Si, and Nb(3 nm)/EuS(20 nm)/Au(20 nm)/Nb(4 nm)/EuS(10 nm)/SiO$_2$//Si structures were prepared by electron-beam evaporation onto thermally oxidised silicon at room temperatures (see **Methods**). The 3-nm-thick top layer of Nb is to protect the structure. The 30-nm-thick EuS is insulating at room temperature with a contact resistance exceeding 10 GΩ (see **Supplementary Fig. S1**), and $d_s$ varies from 2 nm to 20 nm.

We first discuss the superconducting and magnetic properties of Nb(3 nm)/EuS(30 nm)/Nb($d_s$)/SiO$_2$//Si structures. Fig. 1**e** shows the chemistry diagram of Nb (green), Eu (blue), and O (red) derived from the scanning transmission electron microscope (STEM) showing evidence for oxidation of the Nb capping layer. X-ray reflectivity measurements confirm the thickness of each layer (see **Supplementary Fig. S2**). Fig. 1**f** shows $T_c$ vs $d_{Nb}$ for these structures, showing a decay in $T_c$ with relatively large values of $T_c$ of 0.2 K and 2.1 K for only 2- and 3-nm-thick Nb films, respectively. We define $T_c$ as the mid-point of the superconducting transition from a resistance vs temperature ($R(T)$) measurement. The current bias (1-10 µA) used to determine $T_c$ is sufficiently low and had no measurable effect on $T_c$ itself (see **Supplementary Fig. S11**). The width of the superconducting transition, $\sigma_{Tc}$, defined as the difference in temperature between 90% and 10% of the superconducting transition, is plotted in Fig. 1**f** showing relatively sharp transitions.

In Fig. 1**g** and 1**h** we have plotted the in-plane magnetic field trace of $R(H)$ of Nb(3 nm)/EuS(30 nm)/Nb(2 nm)/SiO$_2$//Si and Nb(3 nm)/EuS(30 nm)/Nb(3 nm)/SiO$_2$//Si unpatterned structures at temperatures across $T_c$. These show that near $T_c$ there is a local minimum in $R$ at the magnetic fields matching the coercive field ($H_c$) of the EuS layer, indicating recovery of superconductivity in the demagnetised state of EuS. The magnetisation vs in-plane magnetic field ($M(H)$) hysteresis loop in the top panel of Fig. 1**h** for the 30-nm-thick control sample of EuS shows that $H_c$ is about ±5.5 mT at 1.8 K. We note that the Curie temperature ($T_{Curie}$) of EuS is similar to the bulk value of about 16.6 K (see Ref. 41 and **Supplementary Fig. S3**). The resistance minima in $R(H)$ match the $H_c$ of EuS of ±5.5 mT and are related to the recovery of superconductivity due to a reduction in $h_{ex}$ in Nb in the demagnetised state of EuS[18,21]. In the magnetised (single domain) state, $h_{ex}$ is maximal thus maximising the suppression of $T_c$. The maximum measured shift in $T_c$ between magnetised and demagnetised states of EuS is about 150 mK for both the 2-nm- and 3-nm-thick Nb layers with the shift decreasing to zero as $d_{Nb}$ approaches the measured dirty-limit coherence length value of $\xi_s$ = 4.6 nm (see **Supplementary Fig. S4 and S5**). These results demonstrate a robust magnetic proximity in superconducting Nb on a single layer of EuS.

We now discuss the performance of the superconducting switches. In Fig. 2**a** we have plotted the in-plane $M(H)$ loop for a Nb(3 nm)/EuS(20 nm)/Nb(4 nm)/EuS(10 nm)/SiO$_2$//Si structure at 4.2 K, which shows a differential switching around ±3 mT and ±6 mT, corresponding to different $H_c$ values of the two EuS layers. By sweeping the magnetic field from positive to negative directions, the relative magnetisation-alignment of the EuS layers changes from P to AP at approximately -3 mT. At -6 mT, the magnetisation of the harder EuS layer switches, recovering a P-state. The extended data of the $T$-dependence of the $M(H)$ loops, remanence, and $H_c$ of the two EuS layers are given in **Supplementary Figs. S6 and S7**. The bottom panel of Fig. 2**a** shows the corresponding $R(H)$ in the superconducting transition at 4.2 K: in the P-state, there is a finite resistance in the normal state with superconductivity recovered in the AP-state which translates to an infinite magnetoresistance, confirming a full superconducting switch effect. We define magnetoresistance as $(R_{H=0} - R_{H=Hc})/R_{H=Hc}$. We note that the switching fields in $R(H)$ do not perfectly match the switching fields in $M(H)$, possibly due to a canted surface magnetic moment on EuS, similar to $R(H)$ scans reported in EuS/Al/EuS structures[18].

In Fig. 2**b** we have plotted the zero-field $T$-dependence of $R_{AP}$ and the $T$-dependence of the normalised resistance mismatch between P- and AP-states derived from individual $R(H)$ scans at each temperature i.e., $(R_P(T) - R_{AP}(T))/R_N(T) = \Delta R(T)/R_N(T)$. Selected $R(H)$ scans at temperatures across $T_c$ are shown in **Supplementary Fig. S9**, and zero-field $R_{AP}(T)$ is obtained using the method described in **Supplementary Fig. S10**. From these measurements we obtain a superconducting switch efficiency of $\Delta T_c/T_{c,AP} = 0.3$ in Nb(3 nm)/EuS(20 nm)/Nb(4 nm)/EuS(10 nm)/SiO$_2$//Si. An efficiency of $\Delta T_c/T_{c,AP} = 0.5$ is determined for the same structure in **Supplementary Figs. S8-10** (Noted as Device 3).

In order to investigate boosting $\Delta T_c/T_{c,AP}$ of F/S/F structure by inserting a single N interlayer[39,40], we fabricated a Nb(3 nm)/EuS(20 nm)/Au(20 nm)/Nb(4 nm)/EuS(10 nm)/SiO$_2$//Si structure with the HM layer of Au at one Nb/EuS interface. The top panel of Fig. 2**c** shows the in-plane $M(H)$ loop of the structure at 1.8 K which closely matches the equivalent structure without Au in Fig. 2**a**. From the normalised $R(T)$ (green curve, Fig. 2**d**) we estimate that $T_{c,AP}$ is about 1.86 K. The additional suppression of $T_c$ most likely arises from the proximity of the thin Nb layer with the 20 nm Au layer. Remarkably, in this hybrid structure we observe an infinite magnetoresistance and a normal state resistance in the P-state down to the lowest measurable temperature of 20 mK. The ability to maintain a non-superconducting normal state for P magnetisations down to 20 mK demonstrates absolute switching.

For comparison, in Fig. 3 we have plotted the superconducting switch efficiency $\Delta T_c/T_{c,AP}$ values in this study to equivalent structures in the literature involving transition metal Fs or rare-earth Fs. EuS/Nb based structures show $\Delta T_c/T_{c,AP}$ efficiencies that exceed values measured in equivalent structures including EuS/Al[10–22].

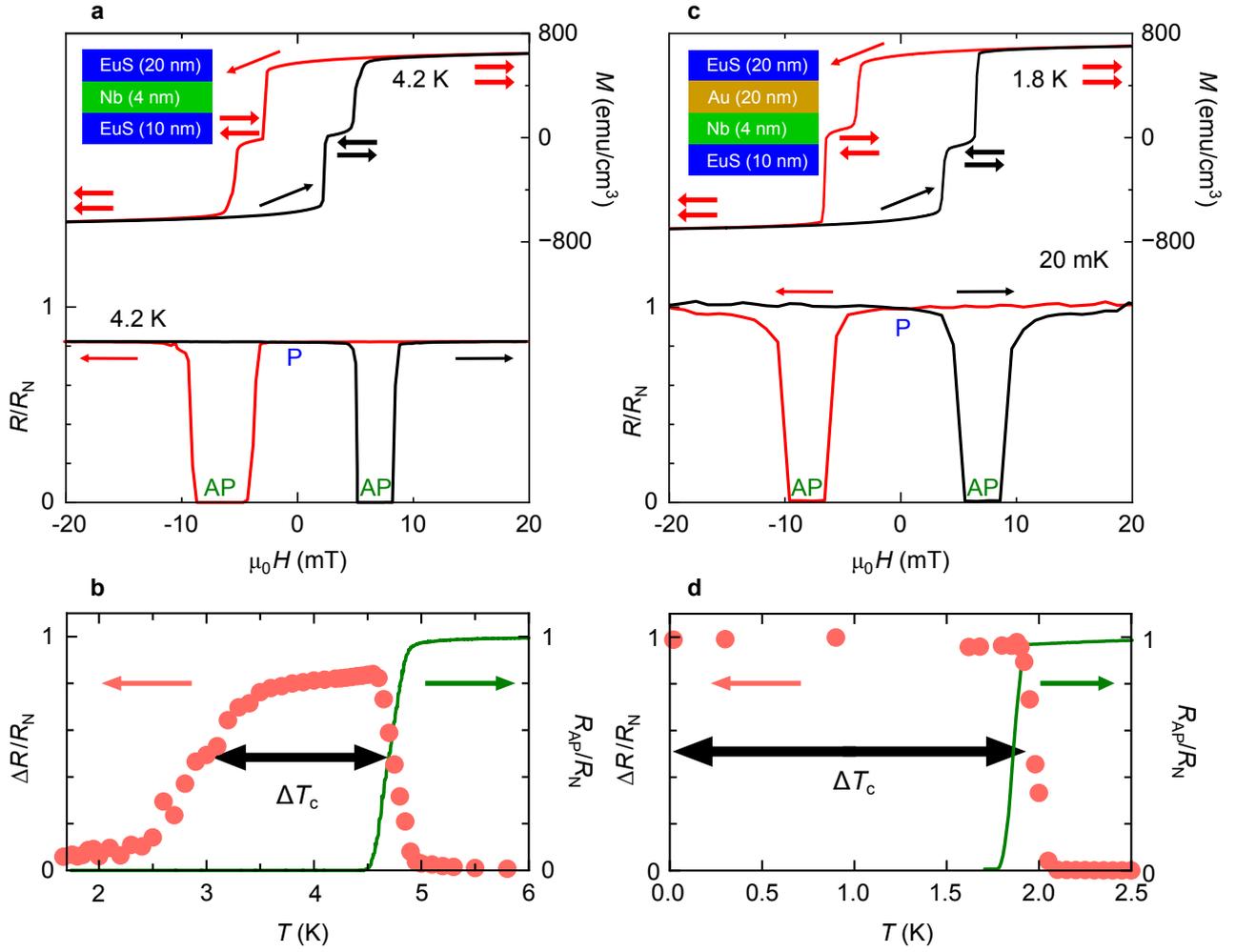

**Fig. 2: Superconducting switch performance with or without a HM interface interlayer. a**, $M(H)$ (right axis) and $R(H)$ (left axis) from an unpatterned Nb(3 nm)/EuS(20 nm)/Nb(4 nm)/EuS(10 nm)/SiO$_2$//Si structure (Device 1) at 4.2 K. Single arrows indicate the magnetic field sweep directions and double arrows represent possible magnetisation directions of the top and bottom EuS layers. Top left inset: schematic cross-section of the structure. **b**, $R_{AP}(T)/R_N(T)$ (green line) and $\Delta R(T)/R_N(T)$ of individual $R(H)$ scans (in pink). **c**, $M(H)$ at 1.8 K (right axis) and $R(H)$ at 20 mK (left axis) of an unpatterned Nb(3 nm)/EuS(20 nm)/Au(20 nm)/Nb(4 nm)/EuS(10 nm)/SiO$_2$//Si structure (Device 2). Top left inset: schematic cross-section of the structure. **d**, $R_{AP}(T)/R_N(T)$ (green line) and $\Delta R(T)/R_N(T)$ of individual $R(H)$ scans (in pink), showing absolute switching with $\Delta T_c/T_c$(AP) equal to 1 (approximately). Data below 1 K are for the same structure measured in a different cooling in a dilution fridge.

The enhancement of $\Delta T_c/T_{c,AP}$ due to the heavy metal layer of Au is, at first glance, unexpected. Firstly, Au has relatively strong spin-orbit coupling, which smears the induced spin splitting of the superconducting density of states in Nb due to the EuS thereby countering the suppression of $T_c$ caused by the proximity-induced magnetic exchange field interaction[45]. Therefore, one would in fact expect a smaller contrast between $T_{c,P}$ and $T_{c,AP}$ in the EuS/Au/Nb/EuS structure. Secondly, theory predicts that the proximity exchange field induced in S ($h_{ex}$) is inversely proportional to the layer thickness (i.e., $h_{ex} = \kappa_{int}/d$)[46–49]. $\kappa_{int}$ is a parameter quantifying the interfacial exchange field related to $G_i$ via $G_i \approx \pi G_0 N_F \kappa_{int}$, where $G_0$ is the conductance quantum, and $N_F$ is the Fermi level density of states per spin[50]. If we assume that $\kappa_{int}$ at the EuS/Nb interface equals to the EuS/Au interface, the addition of Au should suppress the effective exchange interaction by increasing the distance between the EuS layers thereby reducing the value of $\Delta T_c/T_{c,AP}$.

Instead, we see a strong enhancement of $\Delta T_c/T_{c,AP}$. This enhancement likely results from an increase in the exchange coupling at the EuS/Au interface relative to the EuS/Nb interface. This is, in principle, not surprising, since the value of $\kappa_{int}$ is sensitive to microscopic details of the interface, including atomic structure and lattice mismatch[50,51]. Indeed, a large interfacial exchange coupling at the EuS/Au interface has been reported elsewhere[52]. Moreover, the addition of the heavy metal layer Au may partially suppress $T_c$ via the inverse proximity effect, favouring the suppression of superconductivity, and hence reduce the critical field. This may add to the suppression of superconductivity in the P state.

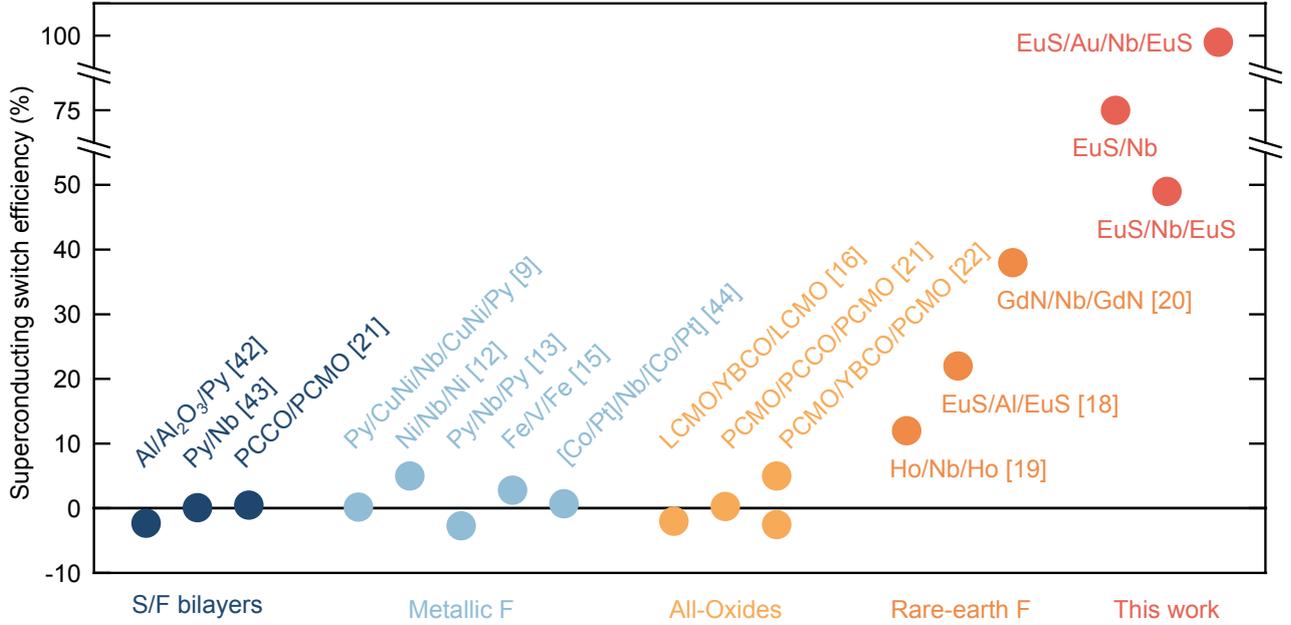

**Fig. 3: Literature survey of superconducting switch efficiencies for F/S/F structures with different materials combinations including transition metal ferromagnets and *f*-orbital ferromagnets.** PCMO is $Pr_{0.8}Ca_{0.2}MnO_3$, PCCO is $Pr_{1.85}Ce_{0.15}CuO_4$, LCMO is $La_{0.7}Ca_{0.3}MnO_3$, and YBCO is $YBa_2Cu_3O_7$.

For a more quantitative understanding, we have calculated the $T_c$ of the different F/N/S/F structures (where F is an insulator) using the Usadel framework based on the quasiclassical Green's functions. Here, we present the main results related to the experiment and provide the details of the model are presented in the **Supplementary Materials**.

In Fig. 4**a** we have plotted the calculated $T_c$ of the EuS/Au/Nb/EuS structure vs Au layer thickness ($d_{Au}$) in the P- (in blue) and AP- (in green) magnetic states. For $d_{Au} \geq 15$ nm, we are able to obtain a complete suppression of $T_{c,P}$ with $T_{c,AP}$ nonzero for an optimised induced exchange coupling with $\kappa_{EuS/Au}$ = 1.5 meV·nm and $\kappa_{EuS/Nb}$ = 1.2 meV·nm, equivalent to $G_i = 2.15 \times 10^{13}\ \Omega^{-1}m^{-2}$ at EuS/Au and $G_i = 1.6 \times 10^{13}\ \Omega^{-1}m^{-2}$ at EuS/Nb interfaces[50]. As expected, the $G_i$ for EuS/Au is larger than for EuS/Nb. Our estimates of $G_i$ are similar to values reported for EuS/Pt ($G_i = 7 \times 10^{12}\ \Omega^{-1}m^{-2}$)[26] in which the EuS and Pt layers are deposited in separate vacuum system, compromising the interface quality which reduces $G_i$. Furthermore, our $G_i$ for EuS/Au is also similar to YIG/Au ($G_i = 1.73 \times 10^{13}\ \Omega^{-1}m^{-2}$)[29].

Fig. 4**b** shows the dependence of the maximum superconducting switch efficiency vs $d_{Au}$. The dashed line is for $d_{Au}$ = 0. This value differs from the $d_{Au}$ ~ 0 nm limit (highlighted by the solid line) due to the finite interface resistance at the Nb/Au interface and the different exchange coupling strengths at the EuS/Au interface. If the Au is thick enough ($d_{Au} \geq 15$ nm), $T_{c,P}$ is suppressed for all temperatures, achieving an absolute superconducting switch.

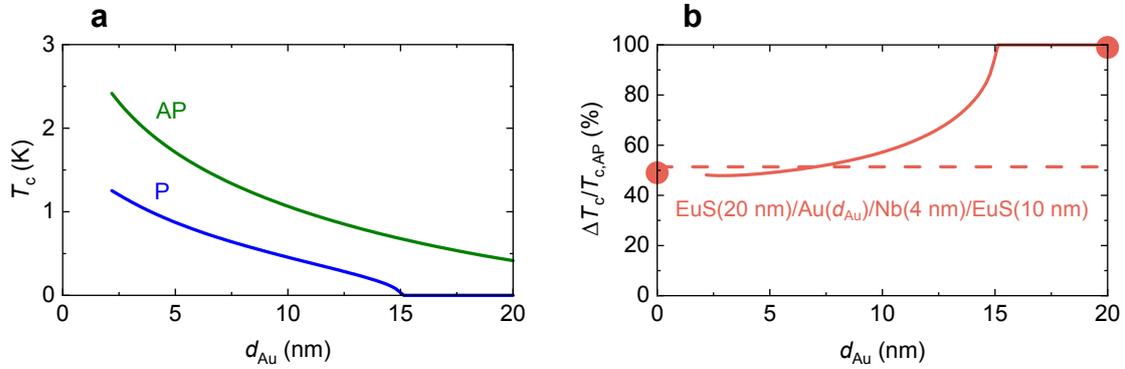

**Fig. 4: Calculated superconducting switch efficiency of EuS/Au($d_{Au}$)/Nb(4)/EuS structures. a**, $T_{c,P}$ (in blue) and $T_{c,AP}$ (in green) as a function of $d_{Au}$. **b**, $\Delta T_c/T_{c,AP}$ as a function of $d_{Au}$. For optimised proximity-induced magnetic exchange fields of $\kappa_{EuS/Au}$ = 1.5 meV·nm at the EuS/Au interface and $\kappa_{EuS/Nb}$ = 1.2 meV·nm at the EuS/Nb interface, absolute switching is expected for $d_{Au} \geq 15$ nm. The dashed line in **b** corresponds to $d_{Au}$ = 0.

In summary, we have demonstrated absolute switching in a EuS/Au/Nb/EuS structure. The switch effect is boosted by the large proximity exchange field induced Au vs Nb which enables absolute on/off switching of superconductivity. The results could create interest in exploiting these effects. For example, a large $\Delta T_c/T_{c,AP}$ ratio is key towards the development of non-volatile superconducting random access memory. Wires which can be controllably switched between superconducting and non-superconducting states are already used in a variety of applications which range in scale from those in persistent mode superconducting magnets, to small scale devices to break SQUID pick-up loops in NMR systems so that large currents are not induced during field ramps, but all current devices are thermally controlled, so that a heater drives the device above $T_c$. A magnetic switch would eliminate the continuous heat load required to hold a thermal switch open (which can be a significant load on the cryogenic system), albeit requiring careful design to eliminate stray field effects for certain applications.

**Methods**

**Film growth:** Thin-films are deposited onto 5 mm × 5 mm area precleaned thermally oxidised silicon substrates at room temperature in a custom-built ultra-high vacuum electron-beam evaporator with a base pressure of 5×10$^{-9}$ mbar. EuS is evaporated directly from EuS powders with an average diameter of less than 44 μm. All materials are evaporated with a growth rate of approximately 1 nm·min$^{-1}$. We investigate Nb(3 nm)/EuS(30 nm)/Nb($d_{Nb}$), Nb(3 nm)/EuS(20 nm)/Nb(4 nm)/EuS(10 nm) (Device 1 and 3), and Nb(3 nm)/EuS(20 nm)/Au(20 nm)/Nb(4 nm)/EuS(10 nm) (Device 2) structures. The 3-nm-thick top layer of Nb is to protect the structure. The different EuS layer thicknesses ensure different coercive fields for independent magnetisation switching between parallel and antiparallel state. The central Nb layer thickness is optimised to be 4 nm which is thinner than the dirty-limit superconducting coherence length of bulk Nb ($\xi_s$) but thick enough to ensure a relatively sharp superconducting transition width.

**Magnetic measurements:** The magnetic moment vs magnetic field measurements are performed in a Quantum Design Magnetic Property Measurement System (MPMS) equipped with a vibrating sample magnetometer superconducting quantum interference device (SQUID). The system can apply up to 7 T using a superconducting magnet with a magnetic moment sensitivity of about 10$^{-8}$ emu.

**Electron Microscopy Characterisation:** Cross-sectional lamellae are prepared using a Dual Beam focused ion beam. Low/high-resolution annular dark field (ADF) imaging and X-ray energy-dispersive spectrum (XEDS) imaging are carried out using an aberration corrected (probe) Thermo Fisher Themis-Z operated at an accelerating voltage of 200 kV. The electron energy-loss spectroscopy (EELS) data are acquired on a Thermo Fisher Titan3 G2 60-300 S/TEM at 300 kV, equipped with a high-brightness field emission electron gun, a monochromator, and a dual-EELS spectrometer. The pixel dwell time is 0.5s with x16 sub-pixel scanning. The EELS data are analysed using Gatan Digital Micrograph software, to obtain elemental quantification from deconvolved and background-removed Nb-M, Eu-N, S-L, and Si-L edges.

**X-ray reflectivity measurements:** Thickness calibration is performed using X-ray reflectometry with a Brucker D8 diffractometer using copper K-α radiation with a wavelength of 1.54 Å. From Kiessig fringes we estimate layer thicknesses using the Leptos software and a genetic algorithm of approximation. The simulation model corresponded to the structure of the original sample and the measurement conditions used in the experiment in each case.

**Superconducting electrical measurements:** Low temperature current-voltage ($I(V)$) measurements are performed using a four-terminal electrical setup. Measurements above 1 K are performed in a cryogen-free system (Cryogenic Ltd) with an in-plane magnetic field and temperature stability of at least 10 mK. Measurements in the mK range are performed in Oxford Instruments Triton 200 Dilution Refrigerator with 6-1-1 vector magnet and 25 mK electron temperature. The *I*(*V*) characteristics are measured using a current-bias of 1-10 µA, and AlSi ultrasonically-bonded contacts on the thin-film multilayers via 4-probe in line.

**Acknowledgements:** Low-temperature transport and volumetric magnetisation measurements were supported by Cambridge Royce facilities grant EP/P024947/1 and Sir Henry Royce Institute recurrent grant EP/R00661X/1. JWAR and HM acknowledges support from the Henry Royce Institute for advanced materials through the Equipment Access Scheme. HM and JWAR acknowledge funding from the EPSRC through International Network and Programme Grants (No. EP/P026311/1 and No. EP/ N017242/1). GY acknowledges funding from the National Key Research and Development Program of China (No. 2022YFA1402600). The electron microscopy work was primarily supported by DARPA under Grant No. D18AP00008. DWM acknowledges support from the Center for Emergent Materials at the Ohio State University, a National Science Foundation Materials Research Science and Engineering Center (Grant No. DMR- 2011876). BW acknowledges support from the Presidential Fellowship of the Ohio State University. Electron microscopy experiments were performed at the Center for Electron Microscopy and Analysis at the Ohio State University. A.H. acknowledges funding from the University of the Basque Country (Project PIF20/05). F.S.B. and A.H. acknowledge financial support from Spanish MCIN/AEI/ 10.13039/501100011033 through project PID2020-114252GB-I00 (SPIRIT) and TED2021-130292B-C42, and the Basque Government through grant IT-1591-22. S.I. is supported by the Academy of Finland Research Fellowship (project No. 355056). S.K. acknowledges funding from the JST FOREST Grant (No. JPMJFR212V).

**Author contributions:** J.W.A.R. had the original idea of the project and developed it with G.Y. and H.M. The samples were prepared by H.M. and G.Y. with help of G.P.M., K.O., and N.S. Electrical transport measurements were carried out by H.M., G.P.M. Y.L., and G.Y. X-ray diffraction was performed by Y.A. The electron microscopy characterisation was performed by B.W. and D.W.M. The model calculation was performed by A.H., S.I., and F.S.B. All authors discussed the results and commented on the manuscript, which was primarily written by H.M., G.Y., and J.W.A.R.

**Supplementary Materials**

**Part I. Extended Data of the Absolute Superconducting Switch**

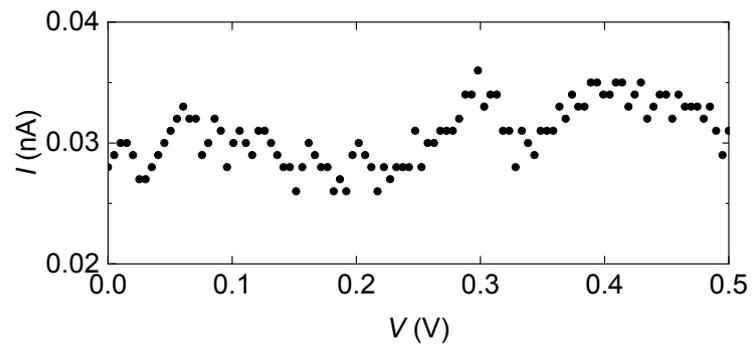

**Supplementary Fig. S1: The current (*I*) - voltage (*V*) characteristics of an uncapped 30-nm-thick EuS thin-film measured at room temperature.** The contact resistance is larger than 10 GΩ ($\rho > 3 \times 10^4$ Ω·cm) In EuS/Nb heterostructures, all current passes through the metallic layer.

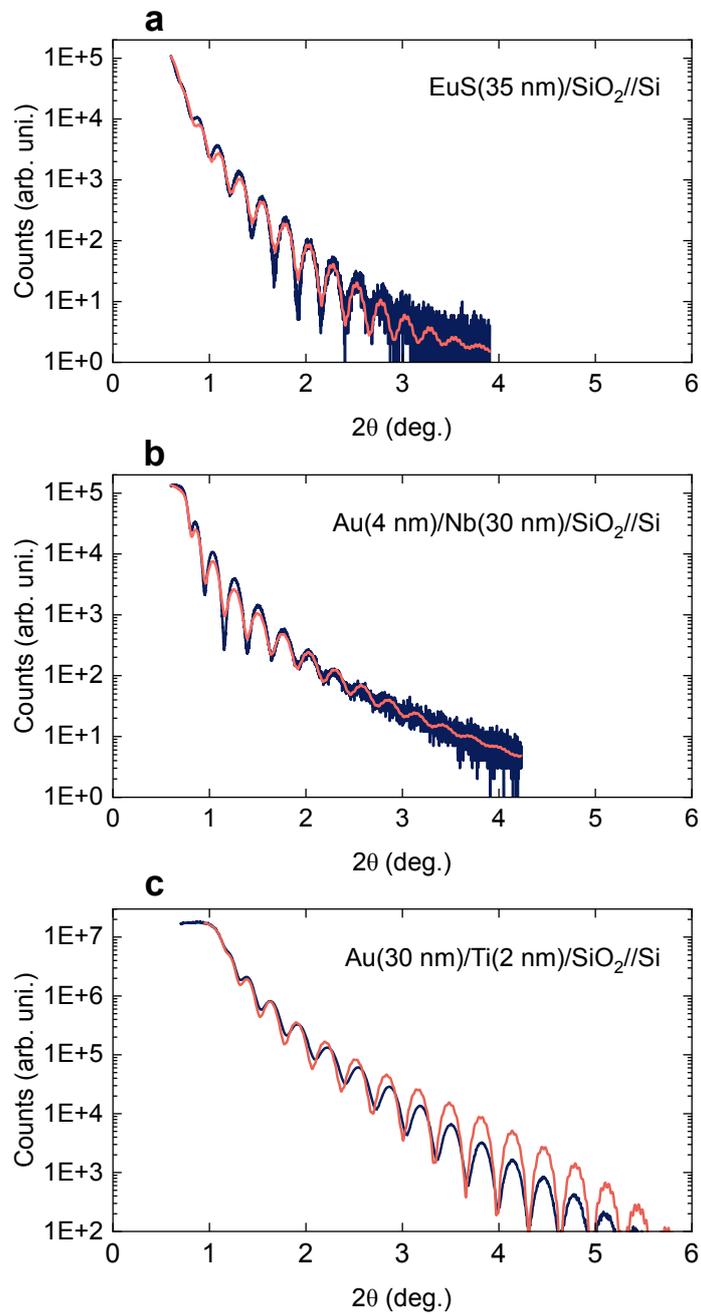

**Supplementary Fig. S2: The X-ray reflectivity measurements.** Top: 35-nm-thick EuS without capping, middle: 30-nm-thick Nb capped by a 4-nm-thick Au layer, and bottom: 30-nm-thick Au with 2-nm-thick Ti seed layer below. Dark blue curves are the measurement data and red curves are the fitting curves.

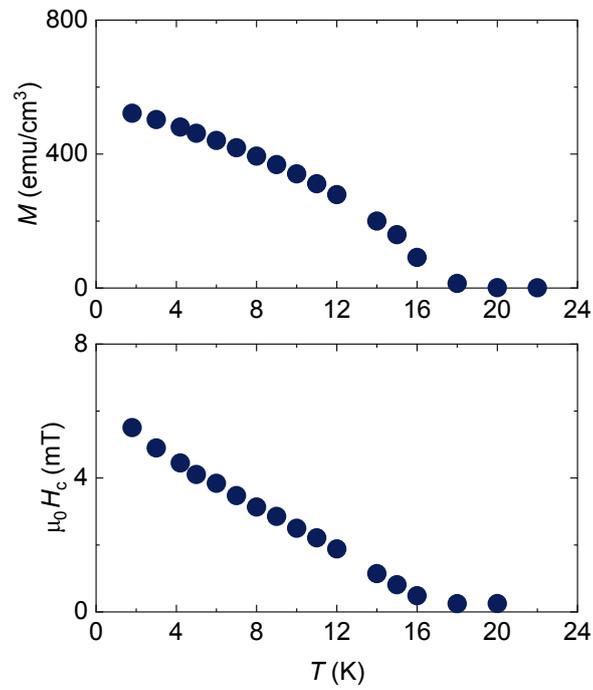

**Supplementary Fig. S3: Remanence (top) and coercive field (bottom) of a 30-nm-thick EuS thin-film as a function of temperature extracted from individual *M*(*H*) loops measured at different *T*.** Curie temperature is close to the theoretical value of bulk EuS of 16.6 K.

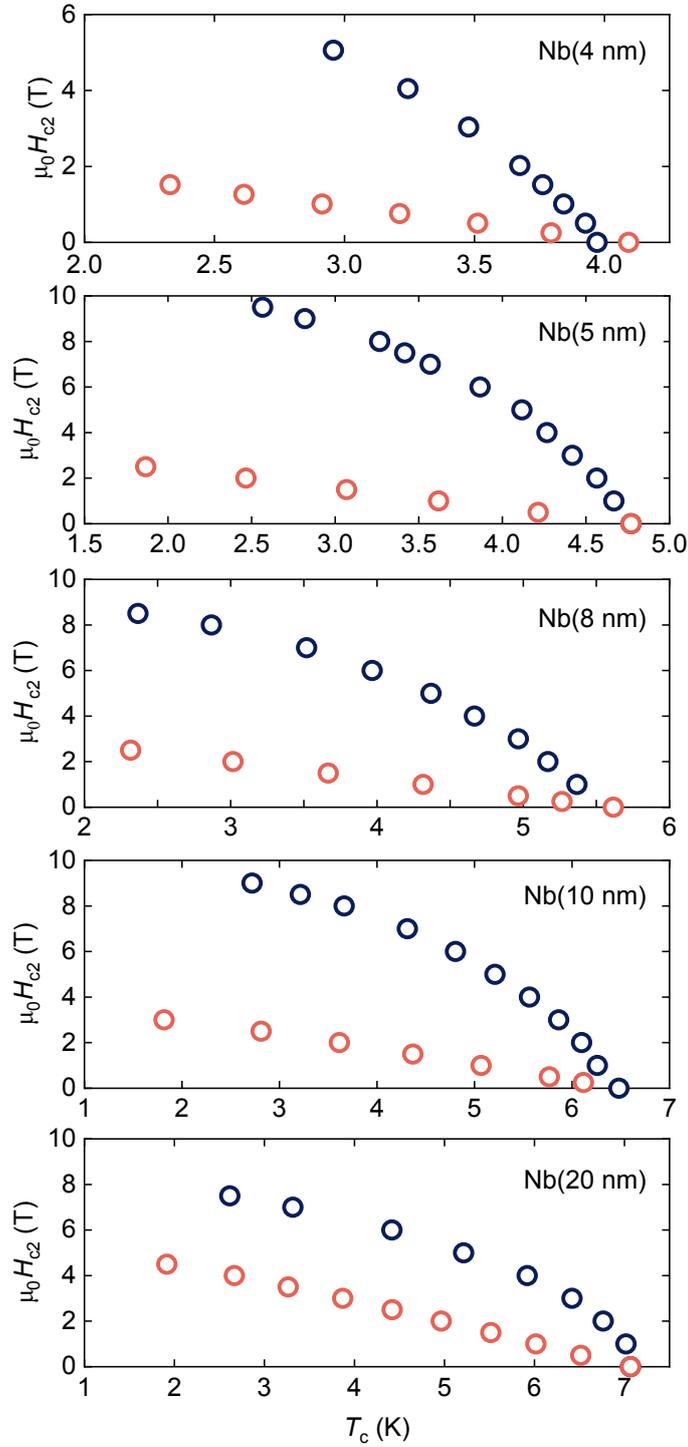

**Supplementary Fig. S4: In-plane (in dark blue) and out-of-plane (in red) critical fields of Nb(3 nm)/EuS(30 nm)/Nb($d_{Nb}$ nm) structures without showing an infinite magnetoresistance.** Nb thicknesses are annotated in the figures. Dirty-limit coherence length ($\xi_s$) of Nb is calculated from the dependence of critical temperature in out-of-plane magnetic fields of the Nb(3 nm)/EuS(30 nm)/Nb(20 nm)/SiO$_2$//Si structure using the relation of $\xi_{GL}(0) = [-(dH_{c2}(T)/dT)(2\pi T_{c0}/\Phi_0)]^{-1/2}$, and $\xi_s = \frac{2}{\pi}\xi_{GL}(0)$, where $T_{c0}$ is the critical temperature at zero magnetic field, $\Phi_0$ is the flux quantum, and $\xi_{GL}(0)$ is the zero-temperature Ginzburg-Landau coherence length. $\xi_s$ of the 20-nm-thick Nb is 4.6 nm.

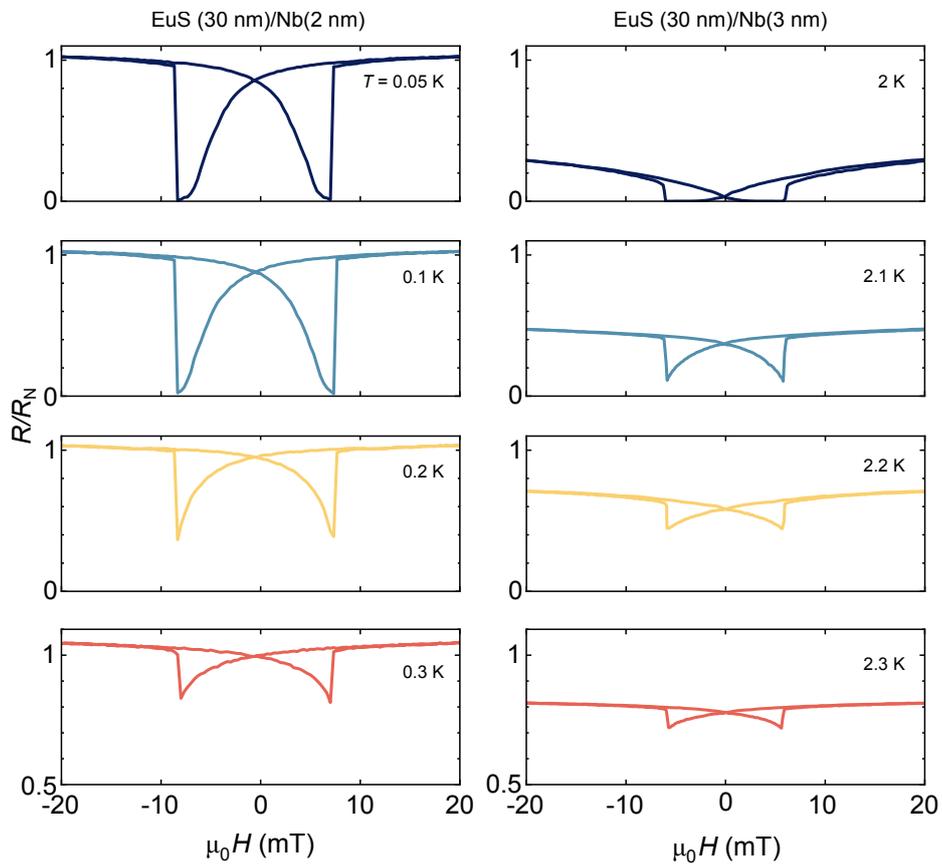

**Supplementary Fig. S5:** Extended data of the normalised $R(H)$ traces of Nb(3 nm)/EuS(30 nm)/Nb(2 nm)/SiO$_2$//Si and Nb(3 nm)/EuS(30 nm)/Nb(3 nm)/SiO$_2$//Si devices across their superconducting transitions.

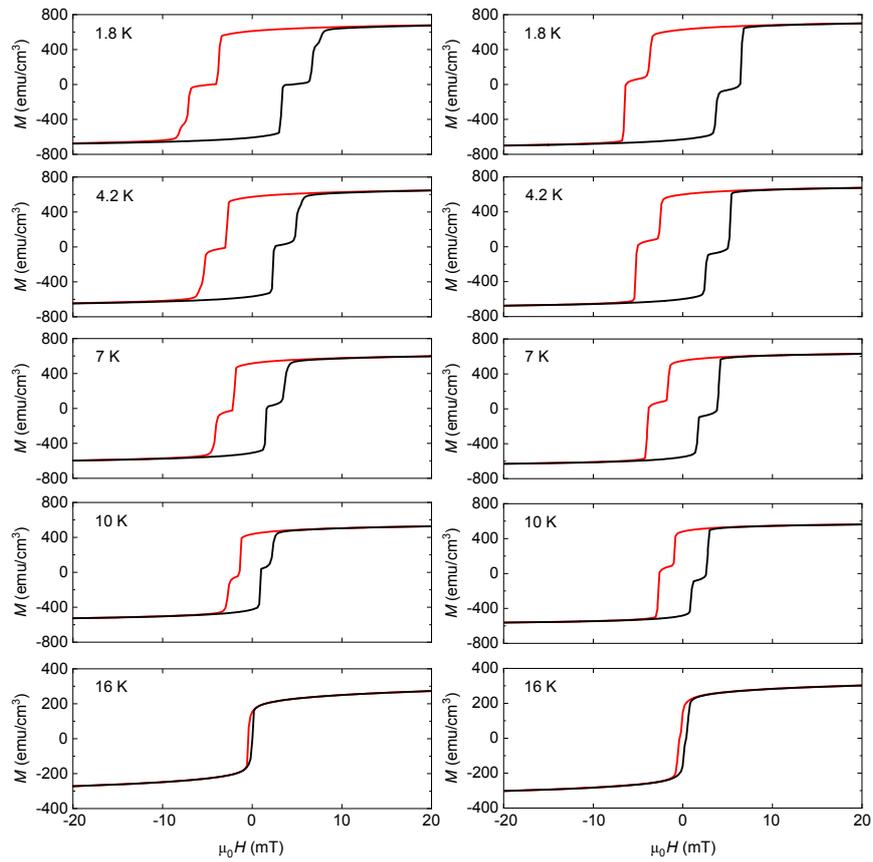

**Supplementary Fig. S6: *M*(*H*) hysteresis loops of Nb(3 nm)/EuS(20 nm)/Nb(4 nm)/EuS(10 nm)/SiO₂//Si (Device 1, left column), and Nb(3 nm)/EuS(20 nm)/Au(20 nm)/Nb(4 nm)/EuS(10 nm)/SiO₂//Si (Device 2, right column).** Red (black) curves indicate a decreasing (increasing) in-plane magnetic field.

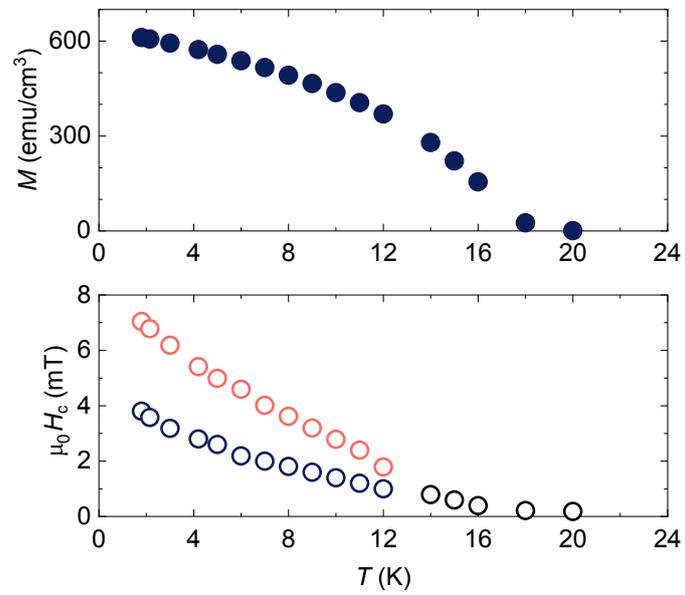

**Supplementary Fig. S7: Remanence (top) and coercive fields (bottom) of a Nb(3 nm)/EuS(20 nm)/Nb(4 nm)/EuS(10 nm)/SiO$_2$//Si structure.** Two distinctive switching steps in the $M(H)$ traces corresponds to the coercive fields of two EuS layers with different thicknesses (in red and blue) is extracted at temperatures below 12 K. At above 12 K, two transition steps in $M(H)$ traces merge into one (in black).

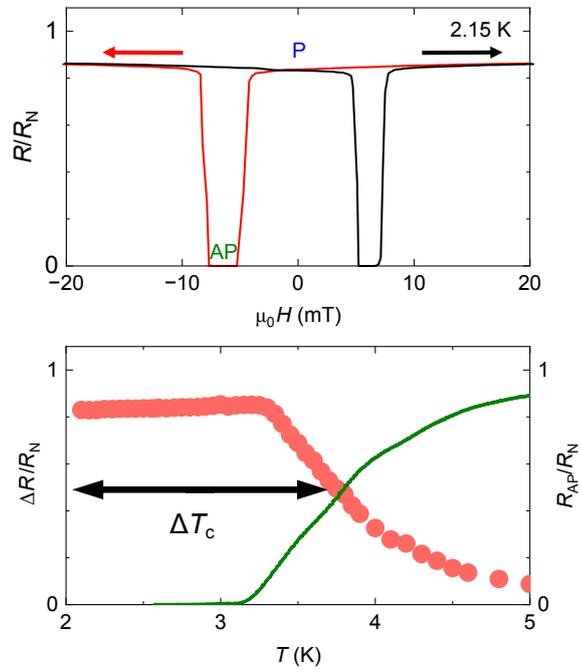

**Supplementary Fig. S8: Superconducting switch performance of an unpatterned Nb(3 nm)/EuS(20 nm)/Nb(4 nm)/EuS(10 nm)/SiO$_2$//Si structure (Noted as Device 2) grown in the same condition as the device shown in Fig. 2a and b.** Top: $R(H)$ at 2.15 K. Red (black) curves indicate a decreasing (increasing) in-plane magnetic field. Bottom: Normalised $R_{AP}(T)/R_N(T)$ (green line, right axis) and $\Delta R(T)/R_N(T)$ of individual $R(H)$ scans (in pink, left axis). $\Delta T_c/T_{c,AP}$ reaches 50 %.

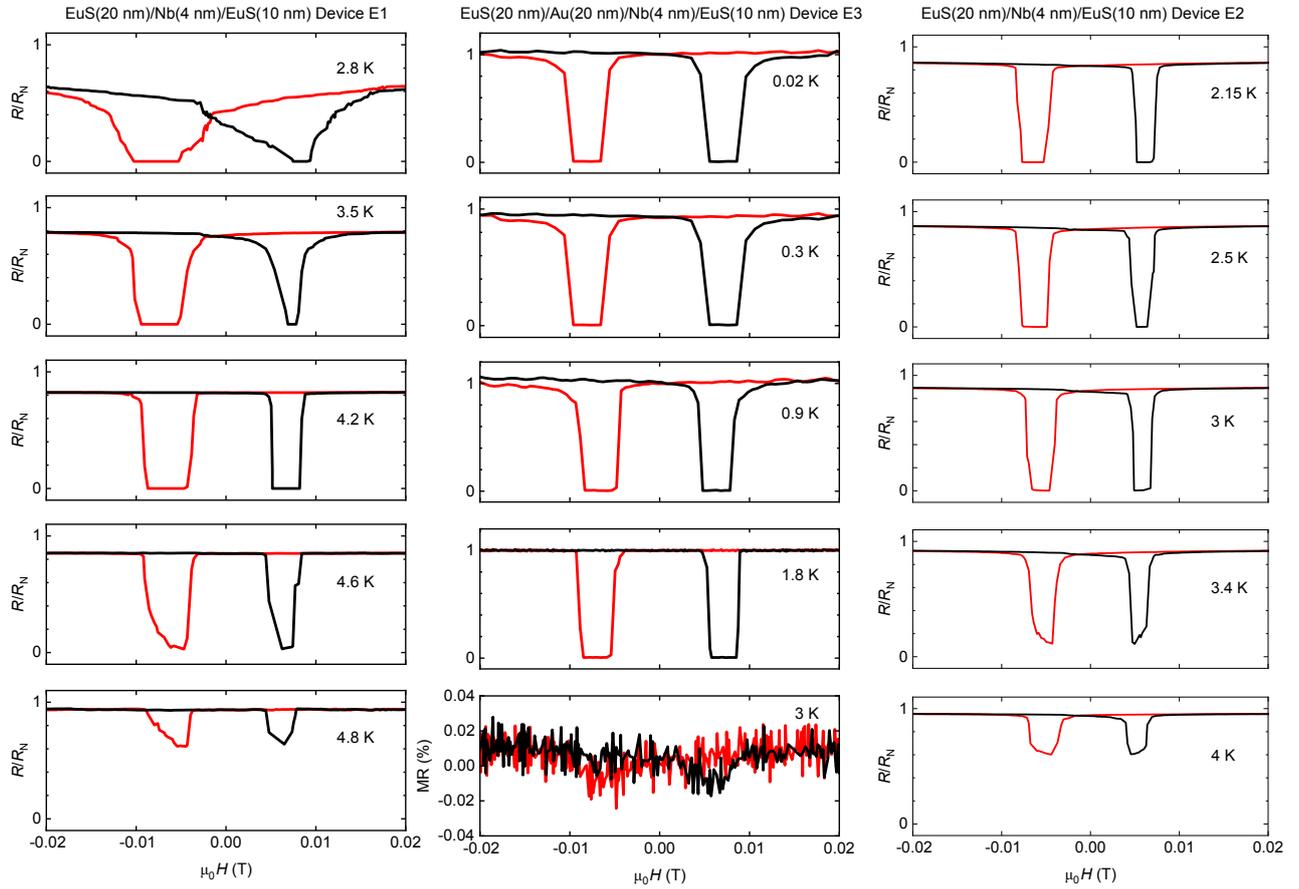

**Supplementary Fig. S9: Extended data of normalised *R*(*H*).** Left column: Nb(3 nm)/EuS(20 nm)/Nb(4 nm)/EuS(10 nm)/SiO$_2$//Si (Device 1), middle column: Nb(3 nm)/EuS(20 nm)/Au(20 nm)/Nb(4 nm)/EuS(10 nm)/SiO$_2$//Si (Device 2), and right column: Nb(3 nm)/EuS(20 nm)/Nb(4 nm)/EuS(10 nm)/SiO$_2$//Si (Device 3), measured at *T* across their superconducting transitions. Red (black) curves indicate a decreasing (increasing) in-plane magnetic field.

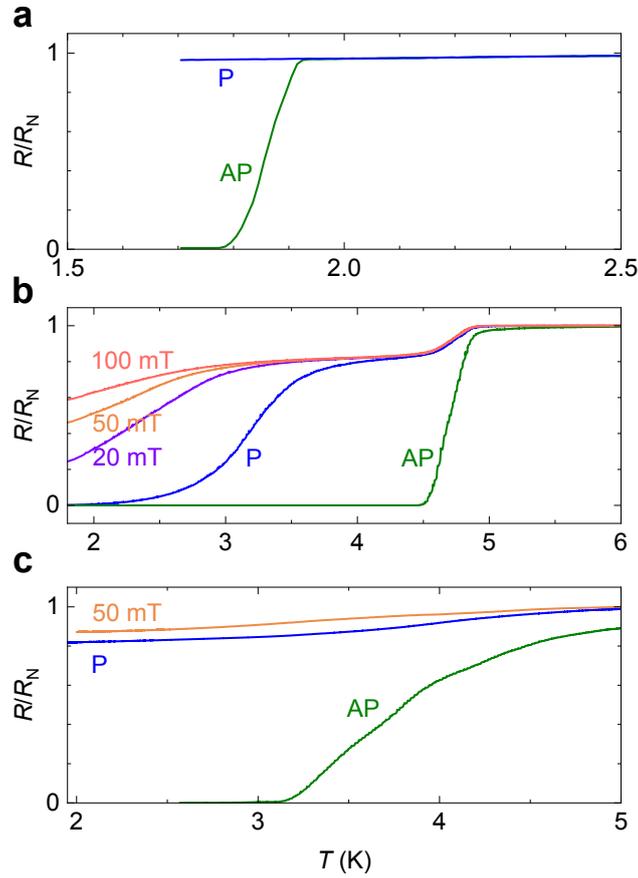

**Supplementary Fig. S10: The temperature dependence of the normalised resistance in the P-, AP-states, and applying in-plane magnetic fields of different switches. a** Nb(3 nm)/EuS(20 nm)/Au(20 nm)/Nb(4 nm)/EuS(10 nm)/SiO$_2$//Si (Device 2). **b** Nb(3 nm)/EuS(20 nm)/Nb(4 nm)/EuS(10 nm)/SiO$_2$//Si (Device 1). **c** Nb(3 nm)/EuS(20 nm)/Nb(4 nm)/EuS(10 nm)/SiO$_2$//Si (Device 3). The zero-field $T$-dependence of $R_P$ and $R_{AP}$ are determined using the following measurement sequence: $R_P(T)$ **trace:** 1: the structure is warmed to the normal state; 2: an in-plane magnetic field of +20 mT is applied to set the P state; 3: the field is then removed so $R_P(T)$ can then be measured in zero-field cooling. $R_{AP}(T)$ **trace:** 1: the device is cooled to the superconducting transition temperature; 2: an in-plane magnetic field of +20 mT is applied to set the P state; 3: an in-plane magnetic field of -4 mT is applied to set the AP state, where the switch is in the zero-resistance state; 4: the field is then removed and the device is warmed to the normal state while maintaining the AP state; 5: $R_{AP}(T)$ is then be measured in zero-field cooling. The other curves are measured in field-cooling with corresponding in-plane magnetic fields. We obtain a superconducting switch efficiency of about $\Delta T_c/T_{c,AP}$ = 30 % in **b**, and larger than 50 % in **c**.

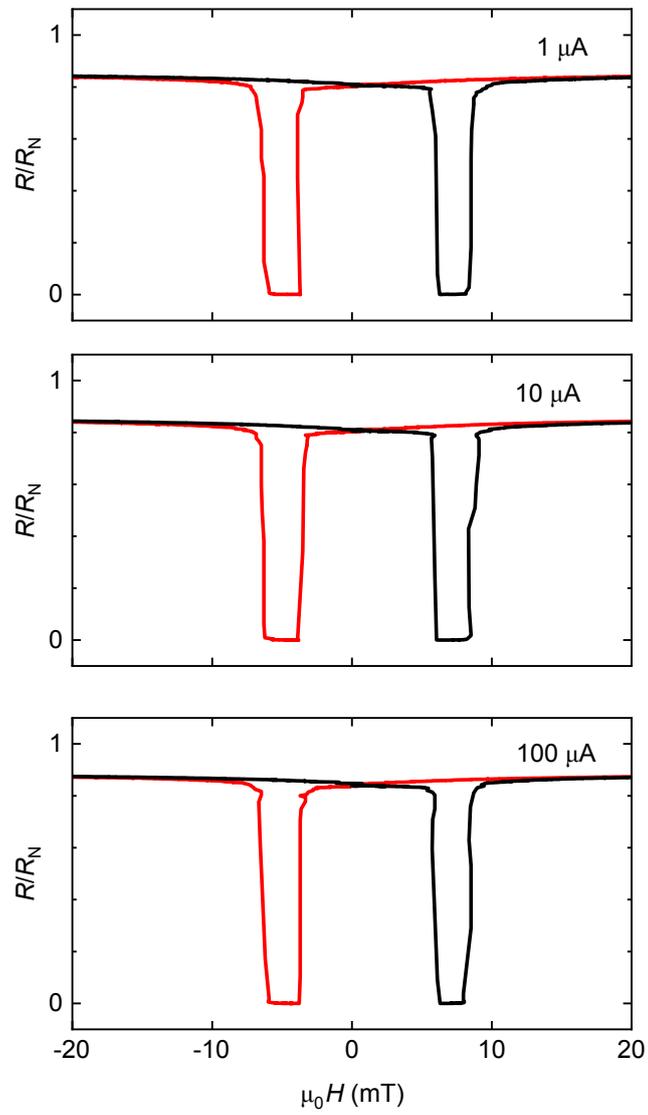

**Supplementary Fig. S11:** *R*(*H*) **of an unpatterned Nb(3 nm)/EuS(20 nm)/Nb(4 nm)/EuS(10 nm)/SiO$_2$//Si (Device 3) structure at 2.15 K with different current bias.** (From top: *I* = 1 µA, 10 µA, 100 µA). There is no significant effect on *R*(*H*). Red (black) curves indicate a decreasing (increasing) in-plane magnetic field.

**Part II. Computing the Critical Temperature of FI/N/S/FI Heterostructures using Quasiclassical Green's Functions**

We study the critical temperature ($T_c$) of the spin-switch using the quasiclassical Green's function (GF) technique[53–58]. In diffusive systems, the quasiclassical GF $\check{g}$ is determined by a diffusion-like equation known as Usadel equation[55]. Together with the normalisation condition $\check{g}^2 = \check{1}$ and the boundary conditions describing the hybrid interfaces determine the value of $\check{g}$, from which the properties of the system, such as the critical temperature, may be extracted.

Assuming that the thicknesses of the layers are much smaller than the coherence length, the GF in the S and N layers can be assumed to be constant, so that the Usadel equation may be integrated over the thickness of the layers. Using the Kuprianov-Lukichev boundary condition[59] to describe the S/N interface, the Usadel equations describing the S and N layers become

$$\left[(\omega + ih\sigma_3)\tau_3 + \Delta\tau_1 + \frac{\sigma_i \check{g}_S \sigma_i}{8\tau_S^{so}} + \Gamma_s \check{g}_N, \check{g}_S\right] = 0, \quad \text{Equation S1(a)}$$

$$\left[(\omega + ih\sigma_3)\tau_3 + \frac{\sigma_i \check{g}_N \sigma_i}{8\tau_N^{so}} + \Gamma_N \check{g}_S, \check{g}_N\right] = 0, \quad \text{Equation S1(b)}$$

where, $\omega = 2\pi T(n + 1/2)$ with $n \in \mathbb{Z}$ is the Matsubara frequency, $\Delta$ is the superconducting order parameter, and $\tau_{so,S/N}$ are the spin-orbit scattering times of the S and N layers, introduced by impurities with spin-orbit coupling. $h_{S/N}$ are the effective exchange fields introduced by the FI layers on the S and N layers. Assuming that the thicknesses of the layers are much smaller than the coherence length, the exchange field may be taken to be homogeneous over each layer, the effective exchange field being inversely proportional to the thickness of the layer $h_{S/N}(d) = \kappa_{int,S/N}/d_{S/N}$[46–49], where $\kappa_{int}$ is a parameter quantifying the interfacial exchange field at the FI/metal interfaces with dimensions of energy times length.

Because we are dealing with superconductivity and spin-independent fields, the GFs on the N and S layers are $4 \times 4$ matrices in Nambu-spin space. The matrices $\sigma_i$ and $\tau_i$ ($i = 1, 2, 3$) in Equation S1 are the Pauli matrices in spin and Nambu space, respectively. Summation over repeated indices is implied. The coupling of the N and S layers is determined by the effective rates[60]

$$\Gamma_S = \frac{v_S}{2\pi d_S \rho_{int}}, \Gamma_N = \frac{v_S^2}{2\pi v_N d_N \rho_{int}}, \quad \text{Equation S2}$$

with $v_{S/N}$ the Fermi velocities and $\rho_{int}$ is a dimensionless parameter describing the resistance of the S/N interface, with $\rho_{int} = \infty$ corresponding to a completely opaque interface. $\Gamma_N$ and $\Gamma_S$ describe the proximity effect and its inverse, respectively.

Close to $T_c$, the GF may be linearised with respect to $\Delta$ as $\check{g} = \text{sgn}(\omega)\tau_3 + \hat{f}\tau_1$, where $\hat{f} = O(\Delta)$ is the anomalous part of the GF, describing the superconducting correlations. The exchange fields introduced by the exchange interaction with the FI layers are either in the parallel or antiparallel configurations, so without any loss of generality we assume that they lie along the z-axis. In this case, the anomalous GF will contain a singlet and a z-triplet projection: $\hat{f}_{S/N} = f_{S/N,0}\sigma_0 + f_{S/N,3}\sigma_3$, with $f_0$ describing the singlet correlations and $f_3$ the triplet correlations. Solving the equation system S1, we obtain the value of the GF at the S and the N layers. The singlet part in the superconducting layer, from which $T_c$ is determined, takes the form

$$f_{S,0} = \Delta \frac{(h_N^2 + \Omega_{N,0}\Omega_{N,3})\Omega_{S,3} - \Gamma_S\Gamma_N\Omega_{N,0}}{\Gamma_S^2\Gamma_N^2 + \Gamma_S\Gamma_N(2h_Sh_N - \Omega_{S,0}\Omega_{N,0} - \Omega_{S,3}\Omega_{N,3}) + (h_S^2 + \Omega_{S,0}\Omega_{S,3})(h_N^2 + \Omega_{N,0}\Omega_{N,3})},$$ Equation S3

where, $\Omega_{S/N,0} = |\omega| + \Gamma_{S/N}$ and $\Omega_{S/N,3} = |\omega| + \Gamma_{S/N} + 1/(2\tau_{S/N}^{so})$.

The critical temperature of the bilayer $T_c$ is given by the self-consistency equation[61]

$$\ln\left(\frac{T_c}{T_c^{BCS}}\right) = 2\pi T_c \sum_{\omega>0}\left[\frac{f_{S,0}}{\Delta} - \frac{1}{\omega}\right],$$ Equation S4

where $T_c^{BCS}$ is the critical temperature of the bulk superconductor. Inserting Equation S3 into Equation S4 and solving $T_c$, we obtain the critical temperature of the bilayer.

We have compared the introduced theoretical model to experimental data to explain the enhancement in the superconducting switch efficiency observed in samples with Au interlayer. We first perform a fitting of the parameters of the model, we consider an Au/Nb bilayer with no EuS layers, i.e. no exchange field $h_S = h_N = 0$. The thickness of Nb of the samples studied was $d_{Nb}$ = 4 nm, while the thickness of the Au layer laid in the $d_{Au} \in [0, 20]$ nm range. $T_c$ of Au($d_{Au}$)/Nb(4 nm)/SiO$_2$//Si bilayers and its theoretical model is shown in **Supplementary Fig. 12**.

The following values for the Au and Nb Fermi velocities $v_{Nb} = v_{Au} = 3 \times 10^5$ m·s$^{-1}$ [62], and the spin-orbit relaxation times $\tau_{Nb}^{so}$ ~ 6 meV$^{-1}$ [63] and $\tau_{Au}^{so}$ ~ 2.4 meV$^{-1}$ [64] were used for the fitting. The Au/Nb interface resistance was extracted from the critical temperature dependence on $d_{Au}$ in Fig. S12, $\rho_{int}$ ~ 20. Next, we consider the EuS/Au/Nb/EuS structure to fit the exchange interaction at the EuS/Nb and EuS/Au interfaces. Appropriate parameters reproducing the enhancement of the efficiency $\Delta T_c/T_{c,AP}$ and absolute switching for thick Au layers $\Delta T_c/T_{c,AP}$ ~ 1 (see Fig. **4a** and **b**) are $\kappa_{EuS/Nb}$ ~ 1.2 meV·nm and $\kappa_{EuS/Au}$ ~ 1.5 meV·nm.

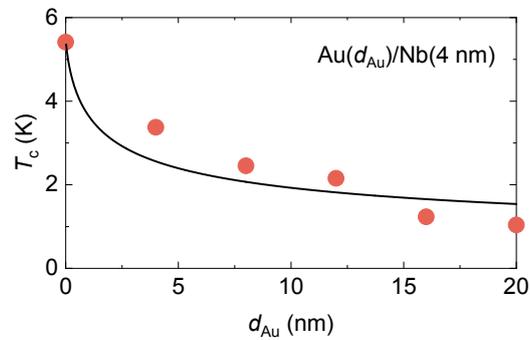

**Supplementary Fig. S12: $T_c$ of Au($d_{Au}$)/Nb(4 nm)/SiO$_2$//Si bilayers.** Nb thickness is fixed to 4 nm. The red data points correspond to the experimental measurements of $T_c$, and the black line is the theoretical model. $T_c$ of 4-nm-thick Nb with $d_{Au}$ = 0 nm is obtained from the $T_c$ of a 4-nm-thick Nb capped by a 2-nm-thick MgO layer.